# Temperature dependent conductivity mechanisms observed in $Pr_2NiTiO_6$

Moumin Rudra*,[1], Saswata Halder[1], Sujoy Saha[2], Alo Dutta[3] and T. P. Sinha[1]

[1] Department of Physics, Bose Institute, Kolkata – 700009, India

[2] Department of Materials Engineering, Indian Institute of Science, Bangalore – 560012, India

[3]Department of Condensed Matter Physics and Material Sciences, S. N. Bose National Centre for Basic Sciences, Kolkata – 700106, India

*Email id: iammoumin@gmail.com

**Abstract**

The crossover between two different conduction mechanisms (variable range and small polaron hoppings) is probed utilizing a conjunction of thermally varying conductivity and impedance in polycrystalline $Pr_2NiTiO_6$ (PNT). A combination of X-ray diffraction (XRD) and Raman spectroscopic investigations authenticates the lower symmetric monoclinic structure for PNT. The conductivity, impedance and relaxation spectra highlight the combined contributions of the electrodes, grain-boundaries and grain microstructures to the carrier dynamics in PNT. The relaxation mechanism has a non-ideal nature with distribution of relaxation times as observed from the dielectric and impedance spectroscopic data. The dc resistivity investigation points towards a transition in the conduction mechanism showing a gradual crossover at temperature 490 K, from small polaron hopping to Mott's variable range hopping due to reduced activation energy.

## 1. Introduction

A recent interest in rare-earth based $RNiO_3$ (with R = rare-earth elements) has emerged due to the observation of novel electrical properties like metal-insulator transition, high temperature superconductivity, etc. [1-3] cause of the tunable bandwidth of the materials which explains variations in localized and delocalized electron patterns. Among the $RNiO_3$, praseodymium nickelate $PrNiO_3$ (PNO) has the highest bandwidth controlled metallicity [4]. PNO is known to be highly unstable thermodynamically and decomposes into more stable $Pr_2NiO_4$ and NiO at high temperatures. Efforts have been made to synthesize PNO under high oxygen partial pressure to stabilize the $Ni^{3+}$ ion. However, the stability of PNO can also be increased by partial occupation of the Ni-site by other noble elements like Ti, Ir [5], Mn [6], Fe [7], and Zr [8]. Although other elements have been used [5-8], Ti has never been used as a stabilizing element for PNO. Thus, in this study, we have doped Ti at the Ni site to achieve a stable double perovskite at ambient condition. A reduced electron delocalization with respect to PNO, due to longer Ni–O–Ti–O–Ni chains compensates for the structural stability of the perovskite arising from the partial reduction of $Ni^{3+}$ in to $Ni^{2+}$.Thus $Pr_2NiTiO_6$ (PNT) represents a $d^8(Ni^{2+})$ – $d^0(Ti^{4+})$ system, which can open supplementary paths to oxide engineering by splitting of the $e_g$ bands through heterostructuring, as reported in recent theoretical approaches to cooperate with the topological insulators [9-13]. The difference in ionic radii $Ni^{2+}$ (0.69 Å) and $Ti^{4+}$ (0.605 Å) in octahedral coordination leads to lower symmetric structure with respect to the orthorhombic *Pbnm* symmetry of the parent materials.

The X-ray diffraction patterns of $Pr_2NiIrO_6$ [5] and $Pr_2NiMnO_6$ [6] confirms that single phase monoclinic $P2_1/n$ structure, whereas the crystal structure of $Pr_2NiZrO_6$ [8] is found monoclinic $P2_1/n$ structure [93.72%] with a secondary pyrochlore phase $Pr_2Zr_2O_7$ [6.28%]. On the other hand, the different hopping models have been suggested to describe the conduction mechanisms for the perovskite oxides in different temperature ranges. Lekshmi *et al.* [6], Mir *et al.*[7] and Mahato *et al.* [8] reported that small polaron hopping (SPH) model is preferable for $Pr_2NiMO_6$ [M = Mn, Fe and Zr] systems. Whereas, other researchers [14-17] reported that SPH model is considered for $RNiO_3$ systems above 500 K. Kumar *et al.* [18] and Chainani *et al.* [19] reported variable range hopping (VRH) in $LaFe_{1-x}Ni_xO_3$ below room temperature. Kumar *et al.* [18] suggested that nickel replaced $Fe^{+3}$ in this series as $Ni^{+3}$, which caused structural distortions due to mismatch between the ionic sizes of $Fe^{+3}$ and $Ni^{+3}$. As a result the unoccupied density of states above the Fermi level shifted towards the Fermi level by 2 eV. They also explained their results on the basis that $LaNiO_3$ has a carrier density of $8 \times 10^{18}$ $cm^{-3}$, while $LaFeO_3$ is an insulator. Substituting Ni in the system is equivalent to carrier doping. This causes an increase in conductivity and a decrease in activation energy. To the best of our knowledge, no attempt has been made to study the crystal structure and the electrical transport properties of the double perovskite oxide PNT.

The electrodes, grain-boundaries and grains are the three main components, which establish the microstructures of the material. In this work, we have employed alternating current impedance spectroscopy (ACIS) in a variable domain of frequency (45 Hz to 5 MHz) and temperature (164 K to 700 K) to investigate the electrical transport characteristics associated with the electrodes, grain-boundaries and grains contribution in PNT.

## 2. Material and Methods

Polycrystalline PNT was synthesized by standard solid state reaction technique. $Pr_2O_3$ (Sigma-Aldrich, 99.9%), $NiCO_3$ (Loba Chemie, 99%) and $TiO_2$ (Laboratory Reagent, 98%) were taken as ingredient materials. Stoichiometries amount of powders commixed in agar mortar in the presence of acetone (Merck) for 8 hrs. at periodic intervals. The slurry was dried, ground well, calcined in an alumina crucible at 1573 K for 14 hrs. and brought to room temperature under controlled cooling at a cooling rate of 1 K/min. Due to poor sinterability of $Pr_2NiTiO_6$, the calcined sample was palletized (thickness, $d$ = 1 mm and diameter, D = 8 mm) using polyvinyl alcohol (Molecular weight ≈ 115000, Loba Chemie) as a binder. Finally, the pellets were sintered at 1623 K for 15 hrs. and cooled down to room temperature by controlling cooling at a cooling rate of 1 K/min.

To characterize the structure of the prepared sample, the room temperature X-ray diffraction (XRD) was carried out with the Rigaku Miniflex II diffractometer having Cu $K_\alpha$ radiation ($\lambda$ = 0.1542 nm). The detective range is $10^0 - 80^0$ with a scanning rate of $0.02^0$ per step. Amassed data were refined utilizing the Rietveld method [20] with the FULLPROF program [21]. In the refinement process of XRD pattern, the background was fitted with the 6-coefficient polynomial function and the peak shapes were described by the pseudo-Voigt functions. Scanning electron microscope (SEM) (FEI Quanta 200) was used in order to capture the microstructural image of the prepared sample. For SEM analysis, the pellet was kept on the aluminum sample holder by a carbon tape with the gold coating on the crack surface of the pellet. The room temperature Raman spectrum of the sample was collected by a LABRAM HR 800 system, which is equipped with a spectrometer of 80 cm focal length, 1800 grooves/mm diffraction grating and a Peltier-cooled charge-coupled device (CCD). A laser of wavelength 488 nm (Ar-ion laser) was used to excite the sample. A 100x objective with NA 0.9 was used to focus the laser beam on the sample. Impedance spectroscopy on the sintered pellet of PNT was performed in the temperature range 164 K to 700 K, utilizing an LCR meter (HIOKI – 3532, Japan). Eurotherm 818p programmable temperature controller was used to control the furnace temperature with an accuracy of ±1 K. This quantifications were performed over the frequency range from 45 Hz to 5 MHz at the oscillating voltage of 1.0 V. Before the experiment, the flat surfaces on both sides of the pellets were cleaned properly and contacts were made from the thin silver paste. The dc resistivity data are collected from I-V characteristic at different temperatures using Keithley (USA) Data Acquisition Multimeter System.

## 3. Result and discussion

### a) Structural analysis

Room temperature XRD pattern of PNT is displayed in Fig.1, showing good agreement between experimental and theoretically simulated patterns. The red circles represent the experimental data and the black solid line represents the calculated diffraction profile obtained by the Rietveld refinement. The green vertical bar denotes the Bragg's positions and the blue solid curve at the bottom represent the difference between the experimental and the calculated patterns. Rietveld refinement authenticates the single phase formation of PNT in monoclinic $P2_1/n$ space group, which is further established by the absence of the characteristic reflections *h0l* with $h$+$l$=2$n$+1 and *0k0* with $k$=2$n$+1 in XRD pattern, withal suggest the space group $P2_1/n$. The superlattice reflection peaks (111) and (311) indicate the presence of rock salt ordering of Ni and Ti ions at the B-site [22-24]. Hence the centrosymmetric space group $P2_1/n$, which sanctions B-sites ordering is adopted here to refine the crystal structure of the PNT. The Rietveld indices were calculated, and the values of the reliability parameters were obtained to be $R_{exp}$ = 7.09, $R_p$ = 5.18, $R_{wp}$ = 6.68 and $\chi^2$ = 0.887. The refined lattice parameters are found to be $a$ = 5.4231 Å, $b$ = 5.4216 Å, $c$ = 7.6829 Å and $\beta$ = 90.116°. A schematic presentation of the PNT unit cell is shown in the inset of Fig. 1 with the distribution of ions at crystallographic Wyckoff positions 4$e$ for $Pr^{3+}$ ions, 2$c$ for $Ni^{2+}$ ions, 2$d$ for $Ti^{4+}$ ions, and 4$e$ for $O^{2-}$ ions. Each $Ni^{2+}$ and $Ti^{4+}$ ions surrounded by six $O^{2-}$ ions constitute the $NiO_6$ and $TiO_6$ octahedra, respectively. The difference between the experimental and calculated results was negligible, indicating the good quality of the refinement. The refined crystallographic parameters are listed in Table 1.

The geometrical tolerance factor ($t$) given by Eqn. (1) [25], serves as an indicator to predict the stability and distortion of the perovskite phase. The dimensionless tolerance factor ($t$) for PNT can be expressed as:

$$t = \frac{r_{Pr}+r_O}{\sqrt{2}\left(\frac{r_{Ni}+r_{Ti}}{2}+r_O\right)} \quad (1)$$

where $r_{Pr}$, $r_{Ni}$, $r_{Ti}$, and $r_O$ are the ionic radii of Pr, Ni, Ti and O ions, respectively. The value of $t$ is found to be 0.89 for PNT compound, which agrees with the predicted value of the monoclinic perovskite phase.

The SEM image of the sintered pellet is shown in Fig. 2. The microstructure of PNT reveals grains of different shapes and sizes. The size of the grains varies from 0.25 μm to 0.7 μm. The density of the pellet is experimentally measured to 6.8 g/cc using Archimedes' principle and was 95% of the theoretical value 7.16 g/cc.

### b) Vibrational Spectroscopy

The Raman spectroscopy profiles for PNT are exhibited in Fig. 3 (a)-(b). The spectrum has ascended contributed from the bands at 83 $cm^{-1}$, 105 $cm^{-1}$, 157 $cm^{-1}$, 204 $cm^{-1}$, 370 $cm^{-1}$, 550 $cm^{-1}$ and 742 $cm^{-1}$ with adscititious comparatively weaker contributions. The spectrum consisting of several phonon modes is fitted utilizing a standard Lorentzian profile. PNT unit cell (Inset of Fig. 1) consists of a three dimensional network of $NiO_6$ and $TiO_6$ octahedra, whereas Pr atom occupies the interstitial position. From Rietveld analysis of XRD pattern, we find the PNT crystallize in a monoclinic $P2_1/n$ structure. The monoclinic $P2_1/n$ symmetry is derived from the cubic prototype $Fm\bar{3}m$ symmetry by the in-phase and anti-phase tilting of the $NiO_6$ and $TiO_6$ octahedra along the [001] direction in the basal plane of the pseudocubic cell corresponding to the Glazer notation $a^-a^-c^+$. It is possible to correlate the obtained Raman active (gerade) modes in both the phases. The correlation between the Raman active modes of both the cubic and monoclinic phases are shown in Table SM-I. From the group-theoretical analysis, the zone center vibrational modes are tenacious in terms of the representation of the $C_{2h}$ point group and the results are listed in Table SM-II. The distortions of the octahedra and the crystal field splitting thoroughly hoists the degeneracy of the $E_g$ and $F_{2g}$ modes of the cubic structure resulting in adscititious Γ phonons becoming Raman active. Due to the 1:1 order characteristics of PNT, which consists of vigorously ($TiO_6$) and impotently ($NiO_6$) bonded octahedra, only the internal modes of the $TiO_6$ octahedron may be considered. In the monoclinic structure, the site group splits all the degenerate modes whereas the crystal field splits each line into two components. The number of atoms in the $P2_1/n$ unit cell is 20 resulting in a total of 60 modes out of which 3 are acoustic, 24 are Raman active and 33 are IR active [26]. Factor group-theoretical analysis of monoclinic $P2_1/n$ symmetry yields 24 Raman active modes as given in Eq. (2).

$$\Gamma^{g}(P2_1/n) = 6T(3A_g + 3B_g) + 6L(3A_g + 3B_g) + 6\nu_5(3A_g + 3B_g) + 4\nu_2(2A_g + 2B_g) + 2\nu_1(A_g + B_g) \quad (2)$$

where $T$ and $L$ represent the translational modes and the librational modes of the Pr-cation, and $\nu_5$, $\nu_2$, and $\nu_1$ represent the bending, antisymmetric stretching and a symmetric stretching mechanism of $TiO_6$ octahedra with the Pr cation at rest. The phonon frequencies and resonance FWHM parameters after deconvolution of the Raman spectra for PNT are listed in Table 2. For a polycrystalline ceramic, the experimental Raman spectrum has a fewer number of modes than prognosticated by group theoretical polarization rules due to the minute correlation field splitting of the samples. However, from correlations among the modes of the cubic $Fm\bar{3}m$ and monoclinic $P2_1/n$ symmetries, we can assign the modes related to the $TiO_6$ octahedra. Fig. 4 shows some representative displacement vectors of the obtained $TiO_6$ octahedra. Taking this into consideration, the breathing vibration ($A_g$ mode) at 742 $cm^{-1}$ assigned to the symmetric stretching mechanism of the oxygen ions along the Ni–O–Ti axis is given by $\nu_1$ peak. The oxygen antisymmetric stretching vibrations assigned to $A_g$ and $B_g$ ($\nu_2$ peaks) modes occur in the range 550 $cm^{-1}$ to 691 $cm^{-1}$. The bending motion of the oxygen anions inside the $TiO_6$ octahedra is characterized by the $\nu_5$ modes at 493 $cm^{-1}$ and 444 $cm^{-1}$. In all the above cases, the Pr-atom was considered to be fixed. The translational modes ($T$) of PNT in which the Pr-cation undergoes a translation is characterized by the frequencies in the range of 83 $cm^{-1}$ to 204 $cm^{-1}$. The rotations of the Pr-cation manifest itself as the librational breathing mechanism ($L$) in the frequency range of 246 $cm^{-1}$ to 370 $cm^{-1}$. The observed phonon frequencies and their consummate group theoretical assignments are given in Table 2.

### c) Conductivity analysis

In order to understand the conduction dynamics and to extract the deterministic parameters influencing the dynamics, the ac conductivity investigation was formulated in a temperature interval of 164 K to 700 K. The frequency variation of ac conductivity at selected temperatures are presented in Fig. 5 (a). An increase in conductivity with frequency complements the presence of trapped charge carriers in the material occurring in a thermally activated dynamical process [27, 28]. The quantified frequency window is characterized by six distinct regions: three plateau and three dispersion domains, perceived due to the effects of electrode-semiconductor interface, grain-boundaries and grains [29]. The value of "$\sigma_{dc}$" can be acquired by extrapolating conductivity value to the zero applied frequency. For room temperature (T ~ 300 K), we have named these three regions as region I, region II and region III (Fig. 5 (b)). In the region I (i.e., $\leq 10^3$ Hz), the plateau represents the total conductivity followed by a dispersion region in which the electrode contribution relaxes. In the region II (i.e., $10^3$ – $10^5$ Hz) the plateau represents the grain boundary contribution to the total conductivity. The grain-boundary contribution relaxes in the dispersion region after this plateau followed by a plateau in region III (i.e., $>10^5$ Hz), which represents the contribution of grains to the total conductivity. With the increase in temperature, the frequency dispersion region gradually moves towards the higher frequency range. The presence of multiple plateau and dispersion regions can be explained by a power law formation with each plateau and dispersion region being explained by a separate power function. In order to fully characterize the conductivity spectra for PNT, a triple power law formalism is adopted given by Eq. (4), highlighting the contributions of electrode-semiconductor interface, grain-boundaries and grains [29].

$$\sigma(\omega) = \sigma_0 + A_1\omega^{k_1} + A_2\omega^{k_2} + A_3\omega^{k_3} \tag{4}$$

Where $\sigma_0$ is the frequency independent conductivity, i.e., dc conductivity ($\sigma_{dc}$), the coefficients $A_1$, $A_2$ and $A_3$ and the exponents $k_1$, $k_2$ and, $k_3$ are the temperature and material geometry-dependent parameters. The experimental conductivity data are fitted by Eq. 4 for the four different temperatures 164 K, 300 K, 500 K and 700 K as shown in Fig. 5 (b). The values of the fitted parameters are listed in Table SM-III.

The value of dc resistivity is calculated from the I-V characteristic at different temperatures, which shows the similar type of behavior as it is calculated from the value $\sigma_0$ of the previous ac conductivity spectra. Fig. 6 (a) shows the thermal variation of dc resistivity for PNT, showing the semiconducting aspect of the material. In Fig. 6 (b), the experimental data are fitted using the small polaron hopping (SPH) model, defined as,

$$\frac{\rho(\mathrm{T})}{\mathrm{T}} = \rho_\alpha \exp\left(\frac{E_a}{k_B T}\right) \tag{5}$$

where $k_B$ is the Boltzmann constant, $\rho_\alpha$ is the pre-exponential factor, $E_a$ is the activation energy, and $T$ is the absolute temperature. The high temperature domain is well fitted by Eq. 5, whereas a severe deviation is obtained in the case of low temperature region below 490 K, showing the presence of two different conduction mechanisms with dissimilar activation energy. The activation energy ($E_a$ = 0.29 eV) extracted from the high temperature region points towards a polaronic conduction of carrier between adjacent neighboring sites [30]. However below 490 K, the non-linear behavior of the experimental observations points towards a conduction dynamics with different origin.

The activation energy ($E_a$) in this region is calculated using the relation $E_a$ = d[ln $\rho$]/d[1/$k_B T$] [31] and the thermal variation of $E_a$ is shown in the inset of Fig. 6(b). The value of $E_a$ decreases with decreasing temperature suggested that at low temperatures, the range of hopping is greater than the distance between the neighboring equivalent sites due to the lower activation energy involved. In this aspect, the experimental data in the low temperature range are fitted with the Mott's variable-range hopping (VRH) formation, which is defined as [32, 33]

$$\rho(T) = \rho_0 exp\left[\frac{T_0}{T}\right]^{\frac{1}{4}} \tag{6}$$

where $\rho_0$ is a constant and $T_0$ is the Mott's characteristic temperature, which can be expressed in terms of the density of states in the vicinity of the Fermi energy, $N(E_F)$, and the localization length "ξ" as follows:

$$T_0 = \frac{18}{k_B N(E_F)\xi^3} \tag{7}$$

Fig. 6 (c) displays the linear fit of $ln\ \rho$ with $1/T^{1/4}$. The parameters $\rho_0$ and $T_0$ are extracted from the linear fit are found to be 9.572 × $10^7$ Ωm and 1789126 K, respectively. The hopping energy $E_h$ (T) for a given temperature T follows the following Eq. [34]

$$E_h(T) = \frac{1}{4} k_B T^{\frac{3}{4}} T_0^{\frac{1}{4}}. \quad (8)$$

In the inset of Fig. 6 (c) shows the thermal variation of the hopping energy $E_h$ (T) and it can be observed that the hopping energy increases from 0.036 eV to 0.082 eV with increase in temperature from 164 K to 490 K, suggesting a polaronic VRH conduction dynamics of the PNT. The mean hopping distance $R_h$ (T) is given by $\frac{3}{8}\xi\left(\frac{T_0}{T}\right)^{\frac{1}{4}}$ [35]. By adopting an estimated value 2.2 Å for ξ [36], the calculated value of $N$ ($E_F$) is found to be 1.09×$10^{22}$ $eV^{-1}cm^{-3}$, which implies a high charge concentration in PNT. The higher value of $N$ ($E_F$) is fortified by many researchers [36]. The hopping distance $R_h$ (T) can also be obtained from the estimated value ξ. The thermal variation of mean hopping distance $R_h$ (T) (Fig. 7 (d)) shows that the hopping distance decreases from 8.431 Å to 6.413 Å with increase in temperature from 164 K to 490 K. It is observed that the minimum $R_h$ (T) at 490 K is of the order of the average distance ($d$) of nickel ions at B sites ($d$ = 6.43 Å), implying that 490 K marks the maximum temperature corresponding to the shortest hopping distance permitted for VRH to occur [37]. Thus dc resistivity data point towards a crossover of small polaron hopping (SPH) to Mott's variable range hopping (VRH) around 490 K.

### d) Impedance analysis

Fig. 7 (a)-(c) show the experimental (points) and fitted (solid lines) complex impedance plane plot (Z-plot) for PNT at different temperatures. At each temperature, the Z-plot shows three semicircular arcs exhibiting the contribution of three different types of relaxation process to the charge transport of the PNT. Due to the non-ideal demeanor of capacitance, the centers of the semicircles are found dejected below the real axis with a spike-like nature in the low frequency region and a diminutive arc in the higher frequency region [28]. The incrementation in diameter of the three semicircular arcs with the decrementation of temperature points towards the thermally activated conduction mechanism in PNT at the electrode, the grain-boundaries and in the grain interiors [38]. In ceramics, the grains are less insulating than grain-boundaries, due to the presence of dangling bonds and non-stoichiometries distribution of oxygen at the grain-boundaries, and can act as carrier traps and form a Schottky barrier for charge transport. The capacitance of this layer is proportional to the reciprocal of the thickness of the layer ($C \propto \frac{1}{d}$, $d$ is the thickness of the layer). The rejoinder of grain-boundaries lies at lower frequencies than that of grains, due to their higher capacitance and resistance [39]. The very high contact capacitance along with more immensely colossal resistance gives the electrode effect at lower frequencies. Consequently, we assign the more minute (at higher frequencies), intermediate and larger (at lower frequencies) arcs to the grains, grain-boundaries and electrode-semiconductor contacts.

In order to understand the contribution of the microstructure to the electrical properties of PNT, an equivalent circuit model connected in series (Fig. 7 (d)) has been employed for fitting impedance data at different temperatures. The constant phase element ($Q$) is utilized for the deviation of capacitance from its ideal demeanor, which may be due to the presence of more than one relaxation processes with approximately kindred or commensurable relaxation times [40]. The capacitance of CPE can be indicated as $C_{CPE} = Q^{1/n}R^{(1-n)/n}$, where $n$ estimates the non-ideal behavior. The value of $n$ is zero for the ideal resistance and unity for the ideal capacitance. Here $R$ and $Q$ are the resistance and constant phase element of grains (g), grain-boundaries (gb) and electrode contact (e) respectively. The solid lines in Fig. 7 (a, b, c) represent the fitting to the electrical equipollent circuit and the fitted parameter $R_g$, $R_{gb}$, $R_e$, $Q_g$, $Q_{gb}$, $Q_e$, $n_g$, $n_{gb}$, and $n_e$ was obtained for four different temperatures 164 K, 300 K, 500 K and 700 K as listed in Table 3. The value of $R_g$, $R_{gb}$, and $R_e$ decreases with incrementing temperature, suggesting the thermal activation of the localized charges [41]. It is observed that $n_e$ and $n_{gb}$ have an increasing proclivity in the quantified temperature range with a value inclining to 1 at and above room temperature. On the other hand, $n_g$ decreases gradually with temperature from 0.95 to 0.92. This suggests that the electrode contact and grain-boundaries capacitance approach the ideal deportment, whereas grain capacitance deviates with increasing temperature. This is due to the existence of the defects such as the release of trapped charges at grain-boundaries and electrode contact with increasing temperature, which is relinquished thoroughly above room temperature with the realization of electronic or ionic defects in the grain interiors.

### e) Dielectric relaxation

Fig.8 (a) shows the frequency dependent dielectric constant (ε′) in the temperature range from 164 to 700 K. These plots show that below 400 K there are only two Debye-like dispersive regions, while above 400 K, two polarization processes are visible and third Debye-like dispersion enters from lower frequency side. With the incrementation of temperature, all these dispersions move towards higher frequencies as one polarization process gradually vanishes from the quantified frequency range on the high frequency side while the other emerges from the low frequency side. Thus, the grain effect is visible at frequencies larger than $10^4$ Hz, while grain boundary relaxation process emerges below $10^5$ Hz in the quantified temperature range, whereas electrode effect dominates at the lower frequency. According to Debye model, all the dipoles follow the applied field and plenarily in the relaxation process below the relaxation frequency of each relaxation process [29]. With the incrementation of the applied frequency, a sudden drop of the ε′ is appeared due to the dipoles lagging abaft the applied field. The value of ε′ becomes approximately independent of the frequency with further increase of the applied frequency, which suggests that most of the dipoles do not respond.

Fig. 8 (c) shows the frequency dependence of tanδ between 164 K and 700 K. It is found that three relaxation peaks are observed in the quantified frequency range, which is due to the presence of electrode, grain-boundaries, and grains. The electrode effect appears on the low frequency side (below 100 Hz). In the quantified temperature range, it is found that all three loss tangent peaks move towards the high frequency side with the incrementation in temperature. Rudimentally, the loss tangent peak depends on the mobility of charge carriers and the temperature. The mobility of the charge carriers increases with temperature and commence to relax at the high frequency side, which results in the mechanism of the loss tangent peak towards higher frequency side. Similarly, the loss peaks for grains are visible above $10^5$ Hz, which shift towards the high frequency side and move out of the quantified frequency range at high temperatures. The tangent loss value of grains is around 0.2 - 0.25 at frequencies above $10^5$ Hz and for grain boundaries, its value is 0.35-0.4 between 45 Hz and $10^5$ Hz depending on temperature. The highest value of tangent loss (0.78-0.8), at lowest frequencies proximate to 45 Hz, is due to the electrode-semiconductor contact. The observed value of tangent loss for grains is cognate to the charge carrier mobility and the number of dipoles available for relaxation process [42, 43]. With the incrementation in temperature, both ε′ dispersion regions (Fig. 8 (a)) and tanδ peaks (Fig. 8 (c)) move towards the high frequency side which suggests the thermally activated deportment of the relaxation process.

The frequency dependence of dielectric constant shown in Fig. 8 (a) can be described by the empirical Cole-Cole equation [44, 45]:

$$\varepsilon^* = \varepsilon' - j\varepsilon'' = \varepsilon_\infty + \frac{\varepsilon_s - \varepsilon_\infty}{1+(j\omega\tau)^{1-\alpha}} \quad (9)$$

where ($\varepsilon_s - \varepsilon_\infty$) is the dielectric strength, τ is the mean relaxation time and α is the degree of the distribution of relaxation time, which varies in between 0 and 1 (α =0 corresponds to standard Debye relaxation). Considering that PNT consists of conducting grains, poorly conducting grain-boundaries and electrode contact, the Cole-Cole expression for ε′ and ε″ are given as

$$\varepsilon' = (\varepsilon_\infty)_g + (\varepsilon_\infty)_{gb} + (\varepsilon_\infty)_e + \frac{((\varepsilon_s)_g-(\varepsilon_\infty)_g)\left\{1+(\omega\tau_g)^{1-\alpha_g}\sin\left(\frac{\alpha_g\pi}{2}\right)\right\}}{1+2(\omega\tau_g)^{1-\alpha}\sin\left(\frac{\alpha_g\pi}{2}\right)+(\omega\tau_g)^{2-2\alpha_g}} + \frac{((\varepsilon_s)_{gb}-(\varepsilon_\infty)_{gb})\left\{1+(\omega\tau_{gb})^{1-\alpha_{gb}}\sin\left(\frac{\alpha_{gb}\pi}{2}\right)\right\}}{1+2(\omega\tau_{gb})^{1-\alpha_{gb}}\sin\left(\frac{\alpha_{gb}\pi}{2}\right)+(\omega\tau_{gb})^{2-2\alpha_{gb}}} + \frac{((\varepsilon_s)_e-(\varepsilon_\infty)_e)\left\{1+(\omega\tau_e)^{1-\alpha_e}\sin\left(\frac{\alpha_e\pi}{2}\right)\right\}}{1+2(\omega\tau_e)^{1-\alpha_e}\sin\left(\frac{\alpha_e\pi}{2}\right)+(\omega\tau_e)^{2-2\alpha_e}} \quad (10)$$

$$\varepsilon'' = \frac{((\varepsilon_s)_g-(\varepsilon_\infty)_g)\left\{1+(\omega\tau_g)^{1-\alpha_g}\cos\left(\frac{\alpha_g\pi}{2}\right)\right\}}{1+2(\omega\tau_g)^{1-\alpha}\sin\left(\frac{\alpha_g\pi}{2}\right)+(\omega\tau_g)^{2-2\alpha_g}} + \frac{((\varepsilon_s)_{gb}-(\varepsilon_\infty)_{gb})\left\{1+(\omega\tau_{gb})^{1-\alpha_{gb}}\cos\left(\frac{\alpha_{gb}\pi}{2}\right)\right\}}{1+2(\omega\tau_{gb})^{1-\alpha_{gb}}\sin\left(\frac{\alpha_{gb}\pi}{2}\right)+(\omega\tau_{gb})^{2-2\alpha_{gb}}} + \frac{((\varepsilon_s)_e-(\varepsilon_\infty)_e)\left\{1+(\omega\tau_e)^{1-\alpha_e}\cos\left(\frac{\alpha_e\pi}{2}\right)\right\}}{1+2(\omega\tau_e)^{1-\alpha_e}\sin\left(\frac{\alpha_e\pi}{2}\right)+(\omega\tau_e)^{2-2\alpha_e}} \quad (11)$$

We have prospered fitted the experimental data for ε′ and tanδ (= ε″/ ε′) using Eqs. (10) and (11) for the temperatures 164 K, 300 K, 500 K, and 700 K as shown in Fig. 8 (b) and 8 (d). The values of the fitting parameters for ε′ and tanδ at different temperatures for PNT are shown in Table 4. A good acquiescent is observed between the experimental data of ε' and tanδ and those calculated from Eqs. (10) and (11). The value of α is zero for an ideal Debye relaxation. Here α > 0, which leads a broader peak than a Debye peak [44, 45]. The relaxation time τ (=1/$\omega_m$) decreases with incrementing temperature, which suggests an incrementation in the dipole density and more expeditious rate of polarization in PNT sample. The value of α's decreases with increasing temperature, designating high dispersive relaxation [46].

The temperature dependence of dielectric constant (ε′) of PNT in the temperature range of 164 – 700 K at different frequencies is shown in the Fig. 9. These curves demonstrate the incrementation in ε′ with temperature and no relaxor-like demeanor is observed. With the incrementation of temperature, the no. of charge carriers increases in the sample, which affects the incrementation in ε′. The dielectric constant shows a step-like increase with increasing temperature, which shifts to higher temperature with increasing frequency. This type of behavior in ε′ is very similar in high dielectric constant materials [47-52]. This type of behavior is caused by an equivalent circuit consisting of the grain contribution, which is connected in series to a parallel RC circuit, where R and C are much higher than the corresponding bulk (grain) quantities. The highly resistive thin layers that generate this RC circuit due to electrode-semiconductor interface and grain boundaries. This effect is more pronounced at lower frequencies, mainly due to the presence of the electrode-semiconductor interface. At lower frequency, the value of dielectric constant is high (~ $10^3$), which can be understood in terms of barrier layer capacitance. A Schottky-type barrier layer exists at electrode-semiconductor contact, which may be composed due to the deviation from defects at the surfaces or different work functions of electrons in the electrode-semiconductor interfaces [53]. This layer distributes a high capacitance by offering aversion to further flow of charges which result in high dielectric constant at lower frequencies and higher temperatures [54, 55].

We have studied the temperature dependence of the relaxation time (τ) corresponding to the peak position in tanδ vs. log ω plot. Temperature dependence of relaxation times (τ) corresponding to the grain-boundaries and grains are plotted in Fig. 10 (a) and (c) by utilizing the Arrhenius equation below.

$$\tau = \tau_o \exp\left(\frac{E_a}{k_B T}\right) \tag{12}$$

here $\tau_o$ is the pre-exponential factor, $k_B$ is the Boltzmann constant, $E_a$ is the activation energy for the relaxation process and $T$ is the absolute temperature. It is observed that the values of $\tau_g$ and $\tau_{gb}$ decrease with incrementing temperature and their temperature dependence follow the Arrhenius law (Eqn. 12) for higher temperatures above 490 K. This is in accordance with our conductivity analysis results in Sec. 3 (d), which also suggests that the relaxation mechanism changes at 490 K. Activation energies corresponding to grain boundary $(E_a)_{gb}$ and grain $(E_a)_g$ calculated above 490 K are found to be 0.268 eV and 0.213 eV, respectively. Fig. 10 (b) and (d) show the evolution of relaxation times ($\tau_g$ and $\tau_{gb}$) with temperature using Mott's variable range hopping (VRH). From the figure, we observed that a good acquiescent between the experimental data and VRH fitting were obtained, which denotes that below 490 K, the carriers are localized by random potential fluctuations and preferred hopping is among the sites lying within the certain range of low energies [56]. These results correlate with the conduction mechanism and dielectric relaxation in PNT.

## 4. Conclusions

In this work, the ac electrical transport properties of polycrystalline PNT synthesized by solid state reaction technique have been explored by the thermally varying conductivity and impedance spectroscopic techniques. A combination of X-ray diffraction (XRD) and Raman spectroscopic investigations authenticates the lower symmetric monoclinic structure for PNT. The $A_g$ mode manifests itself as a breathing vibration of the $TiO_6$ octahedra. The conductivity and relaxation spectra highlight the combined contributions of the electrodes, grain-boundaries and grain microstructures to the carrier dynamics in PNT. The relaxation mechanism has a non-ideal nature with distribution of relaxation times as observed from the dielectric and impedance spectroscopic data. The dc resistivity investigation points towards a transition in the conduction mechanism showing a gradual crossover at temperature 490 K, from small polaron hopping to Mott's variable range hopping due to reduced activation energy. The dielectric relaxation mechanism is analyzed on the basis of the Cole-Cole model. The conduction mechanism is shown to be highly correlated to the relaxation mechanism.

## Acknowledgement

Moumin Rudra acknowledges the financial support provided by the University Grants Commission (UGC), New Delhi in the form of NET JRF (ID no. 522407) with award letter no. 2121551156. Sujoy Saha gratefully acknowledges UGC, New Delhi for providing financial assistance in form of Dr. D. S. Kothari Postdoctoral Fellowship with award letter no. F.4-2/2006 (BSR)/PH/16-17/0026. Alo Dutta thanks to Department of Science and Technology of India for providing the financial support through DST Fast Track Project under Grant No. SR/FTP/PS-175/2013.

# Temperature dependent conductivity mechanisms observed in $Pr_2NiTiO_6$

***Figures:***

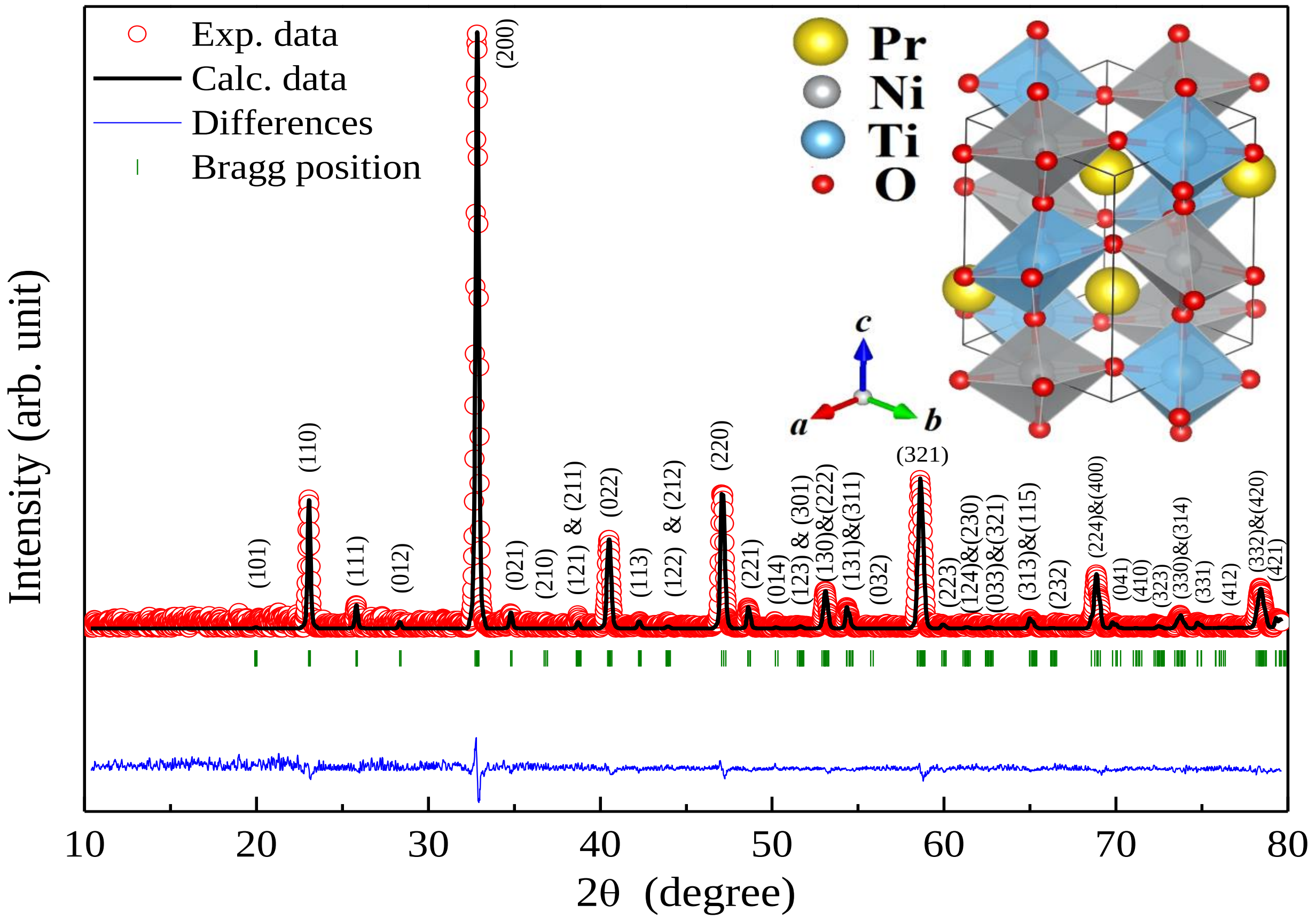

**Fig. 1**

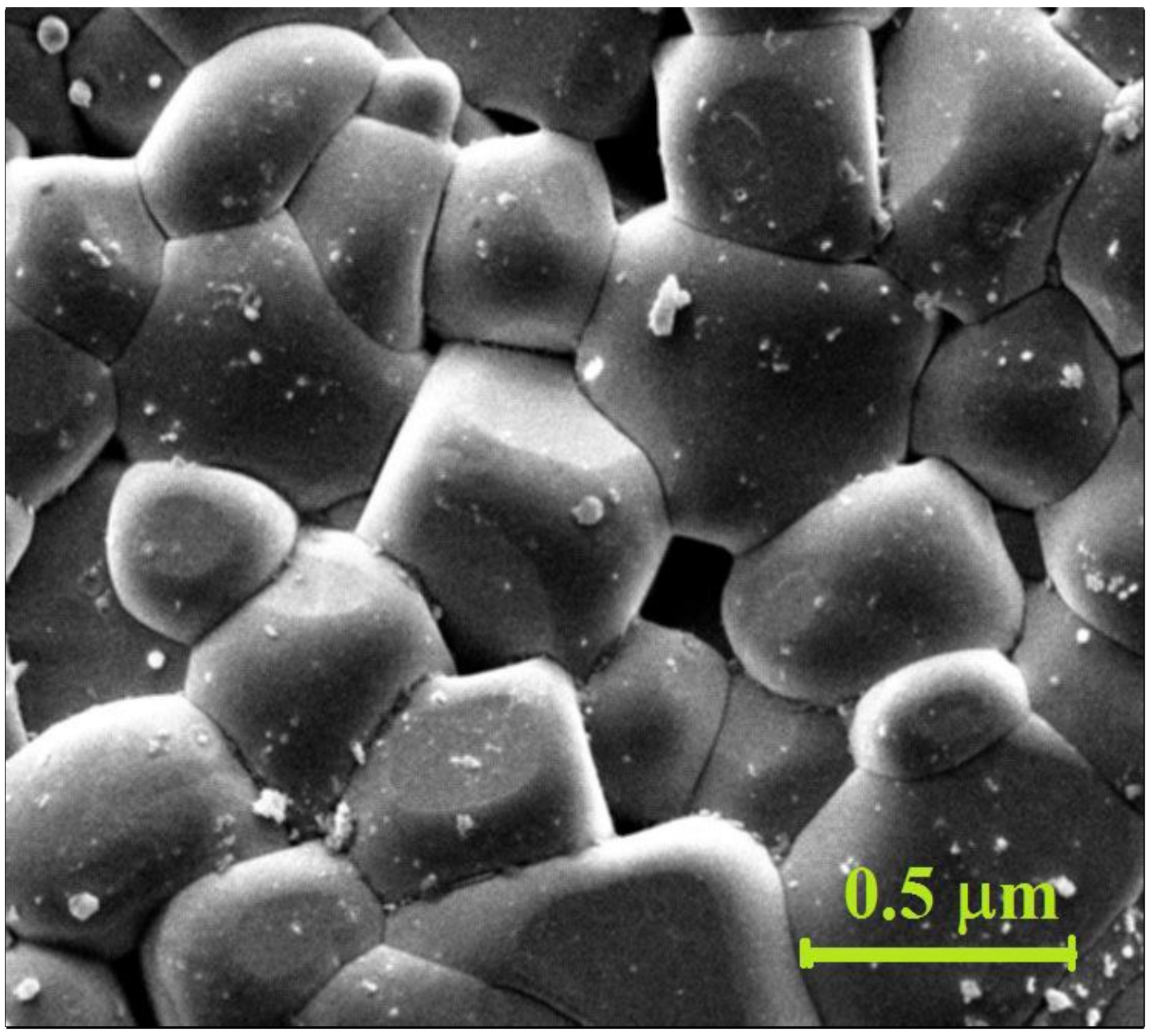

**Fig. 2**

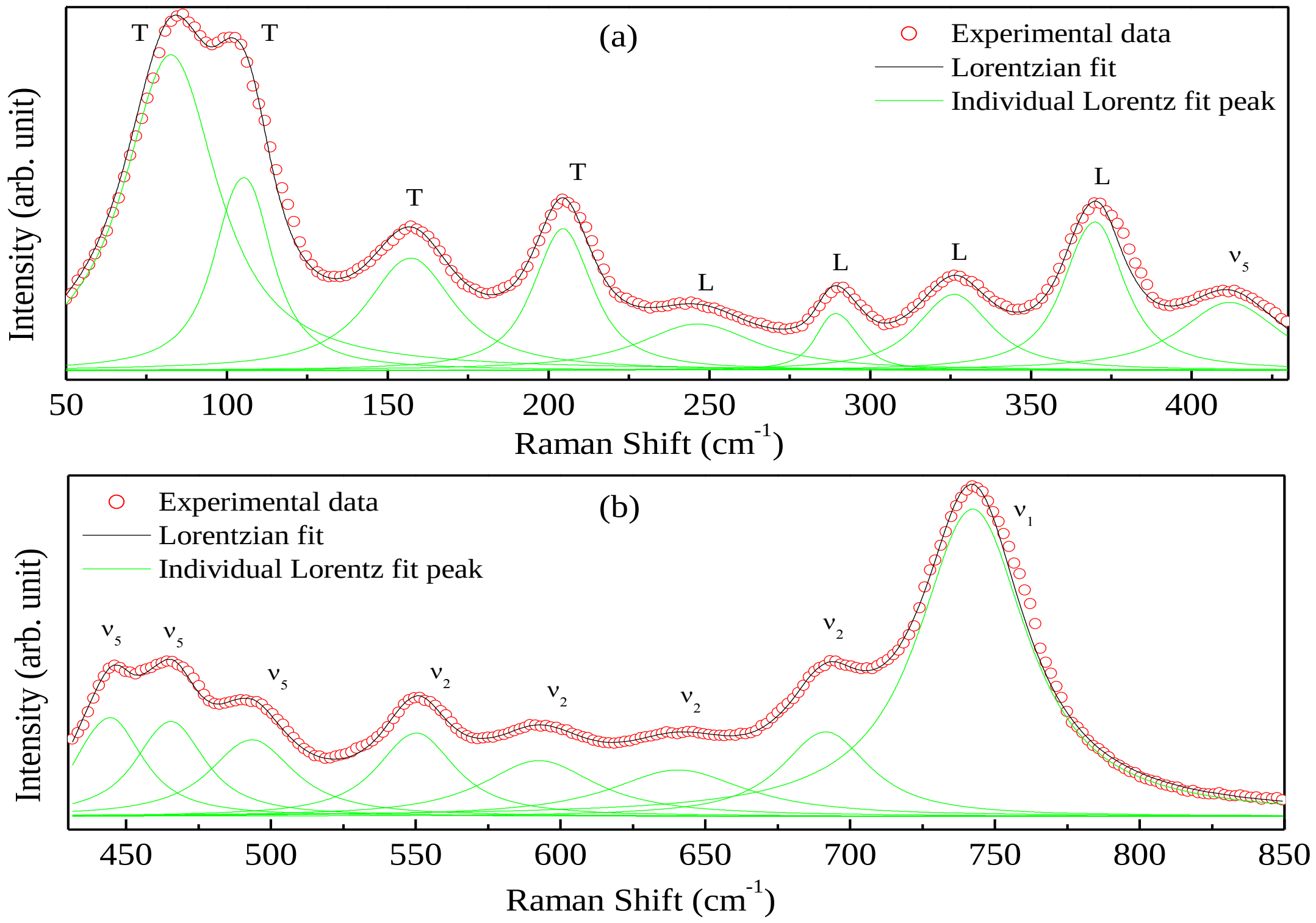
(a)
Experimental data
Lorentzian fit
Individual Lorentz fit peak
T
T
T
T
L
L
L
L
ν5
Intensity (arb. unit)
Raman Shift (cm-1)
50
100
150
200
250
300
350
400
(b)
Experimental data
Lorentzian fit
Individual Lorentz fit peak
ν5
ν5
ν5
ν2
ν2
ν2
ν2
ν1
Intensity (arb. unit)
Raman Shift (cm-1)
450
500
550
600
650
700
750
800
850

**Fig. 3**

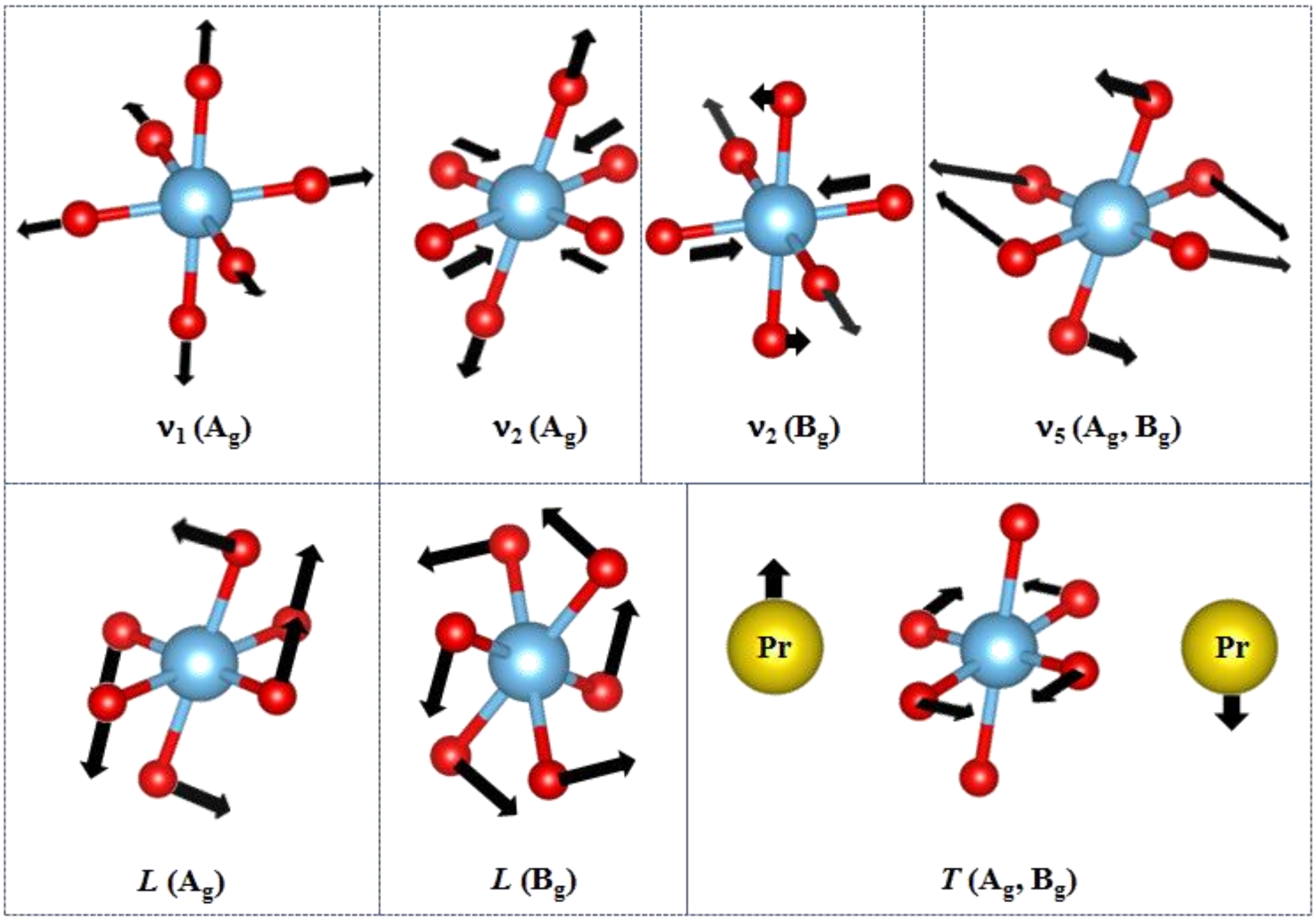

**Fig. 4**

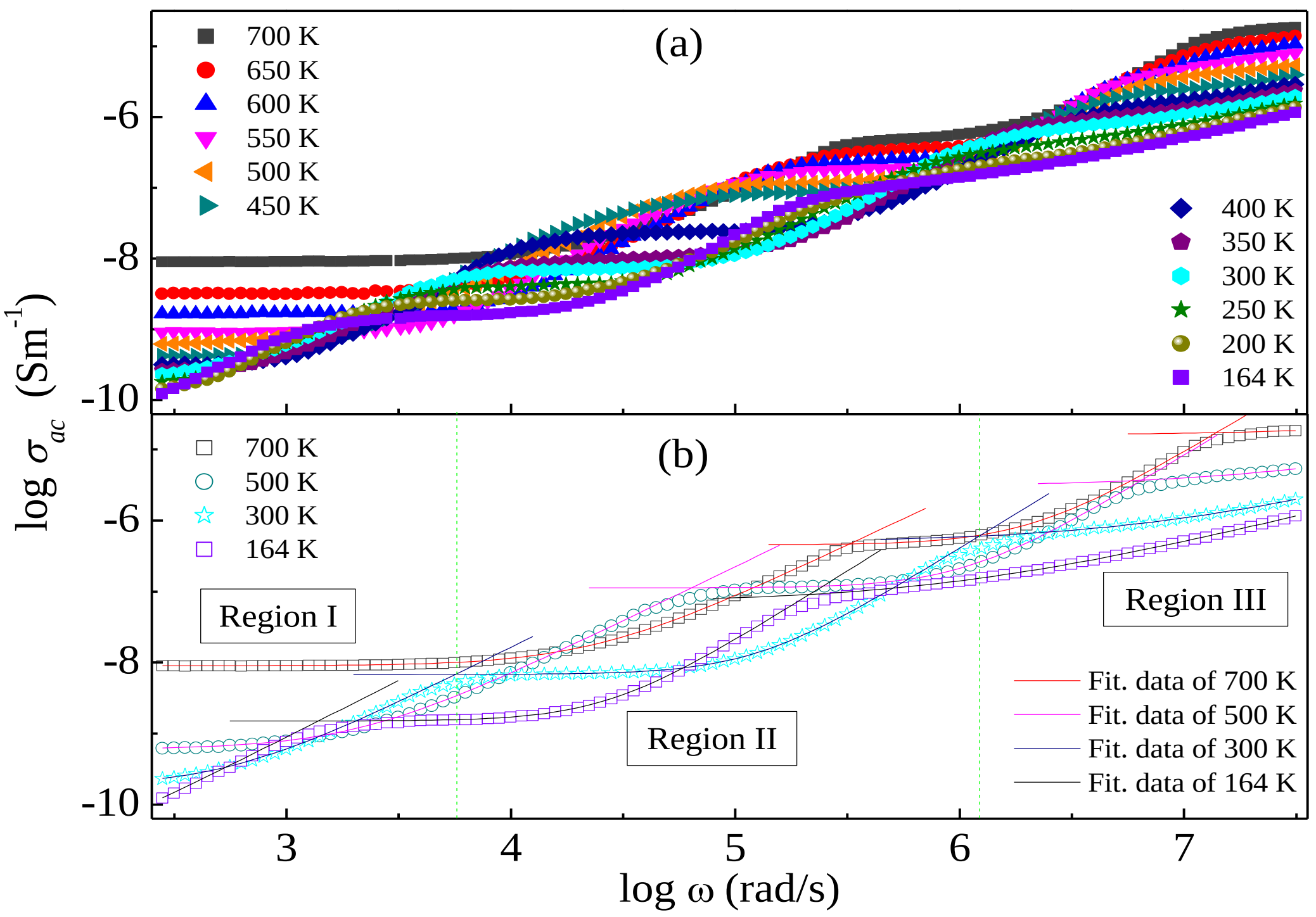
(a)
700 K
650 K
600 K
550 K
500 K
450 K
400 K
350 K
300 K
250 K
200 K
164 K
(b)
700 K
500 K
300 K
164 K
Region I
Region II
Region III
Fit. data of 700 K
Fit. data of 500 K
Fit. data of 300 K
Fit. data of 164 K
-6
-8
-10
3
4
5
6
7
log σac (Sm-1)
log ω (rad/s)

**Fig. 5**

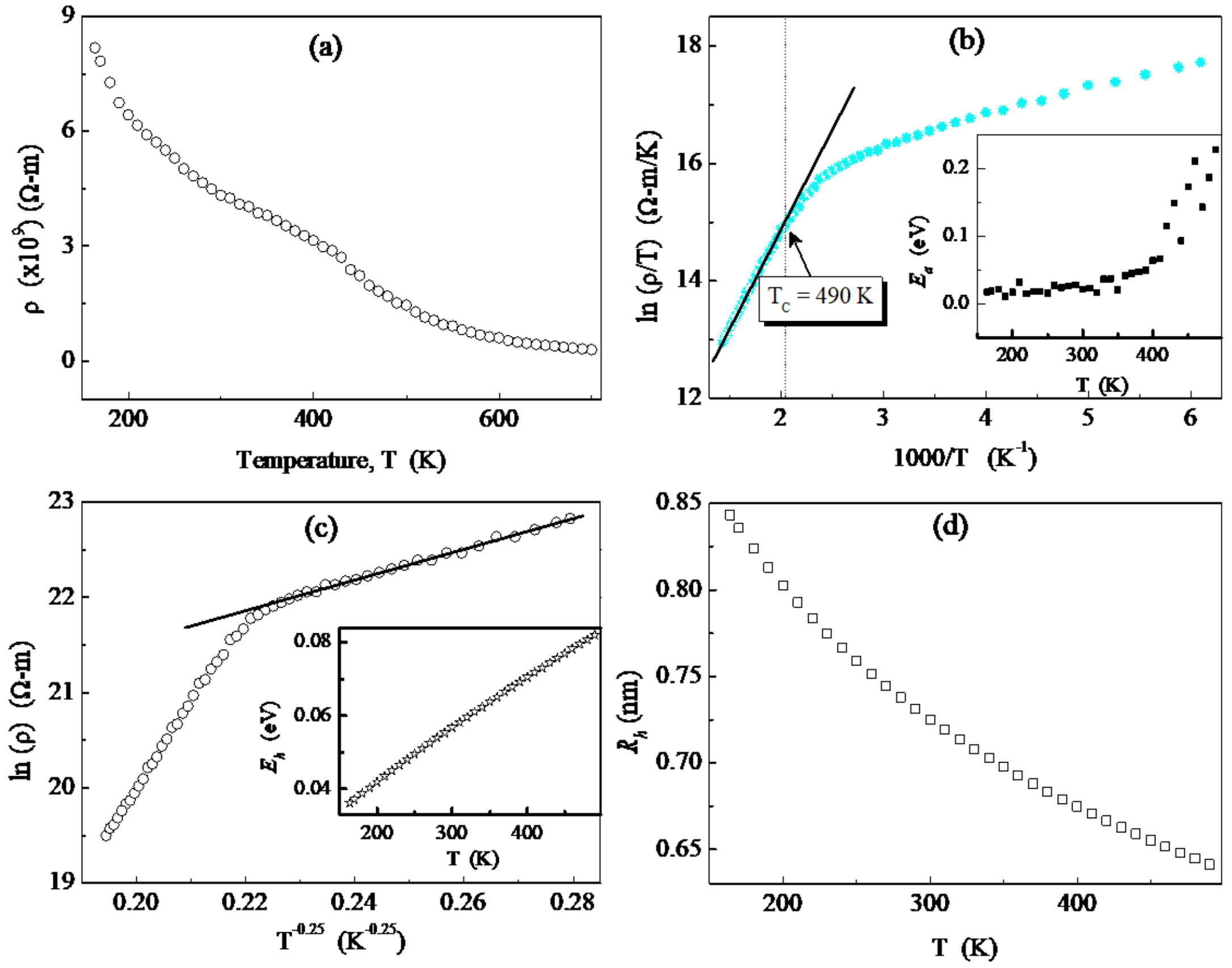
(a)
ρ (x10^5) (Ω-m)
Temperature, T (K)
(b)
ln (ρ/T) (Ω-m/K)
1000/T (K^-1)
T_C = 490 K
E_a (eV)
T (K)
(c)
ln (ρ) (Ω-m)
T^-0.25 (K^-0.25)
E_h (eV)
T (K)
(d)
R_h (nm)
T (K)

**Fig. 6**

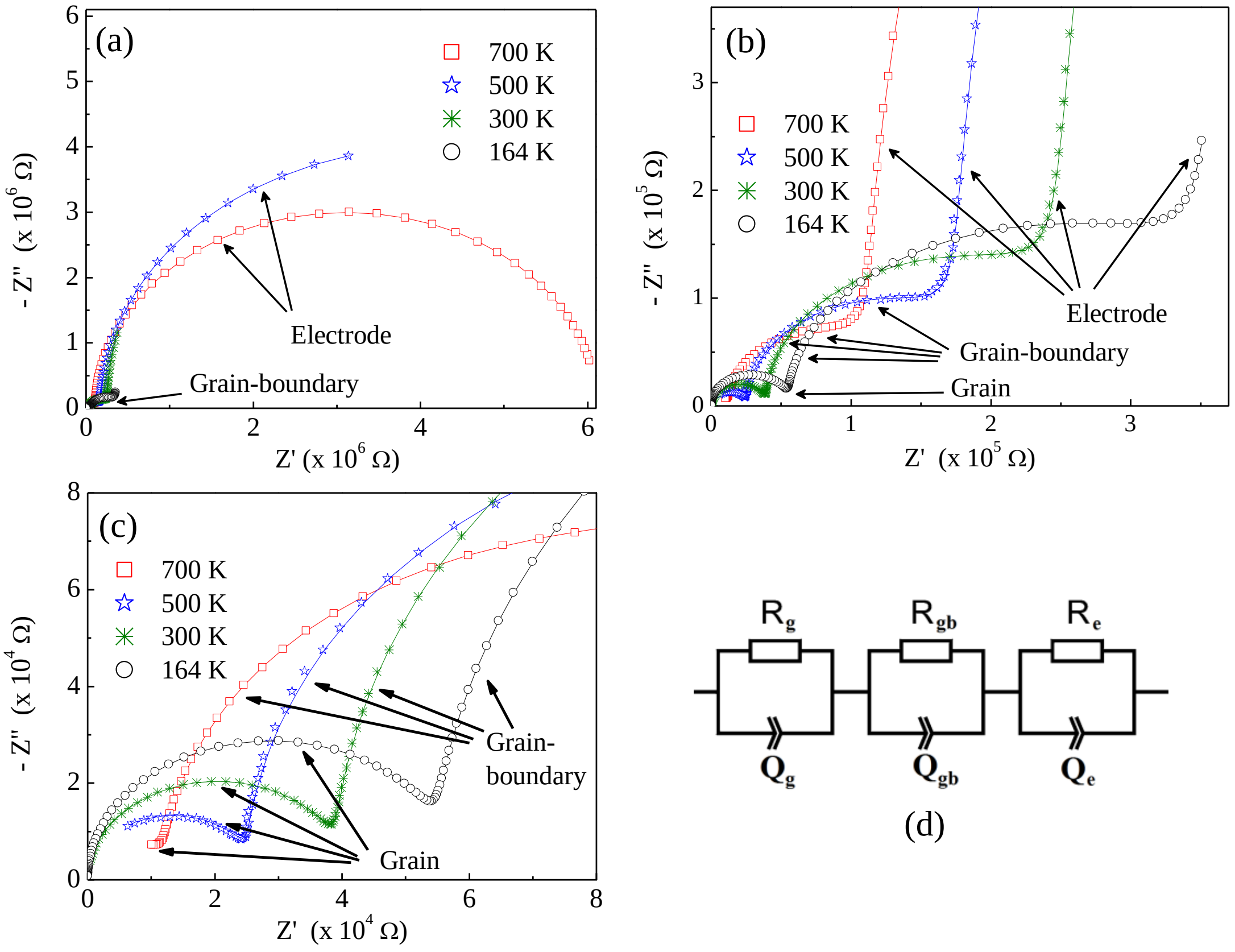

(a)
700 K
500 K
300 K
164 K
Electrode
Grain-boundary
- Z'' (x 10^6 Ω)
Z' (x 10^6 Ω)
(b)
700 K
500 K
300 K
164 K
Electrode
Grain-boundary
Grain
- Z'' (x 10^5 Ω)
Z' (x 10^5 Ω)
(c)
700 K
500 K
300 K
164 K
Grain-
boundary
Grain
- Z'' (x 10^4 Ω)
Z' (x 10^4 Ω)
Rg
Rgb
Re
Qg
Qgb
Qe
(d)

**Fig. 7**

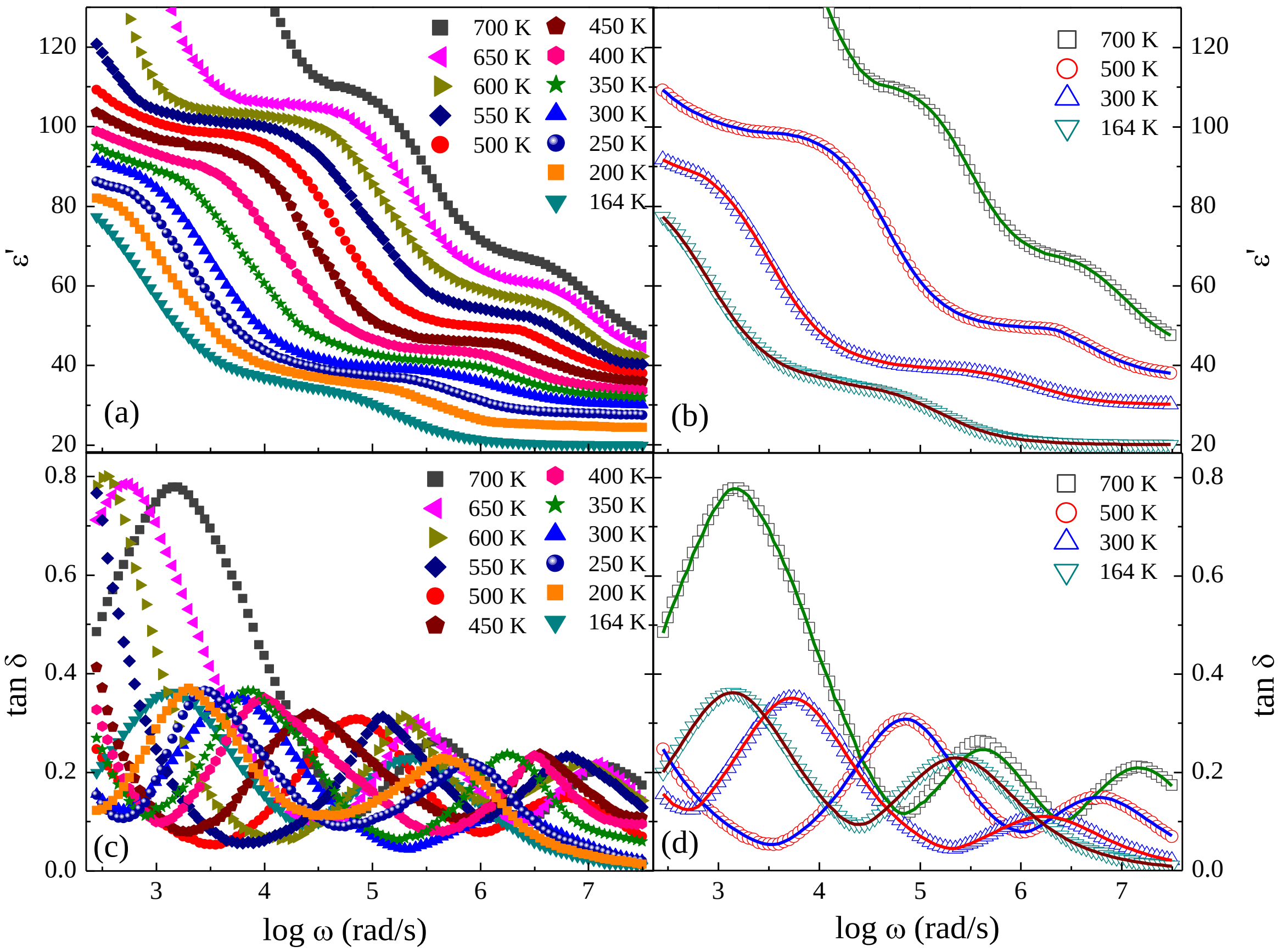

(a)
(b)
(c)
(d)
ε'
tan δ
log ω (rad/s)
700 K
650 K
600 K
550 K
500 K
450 K
400 K
350 K
300 K
250 K
200 K
164 K

**Fig. 8**

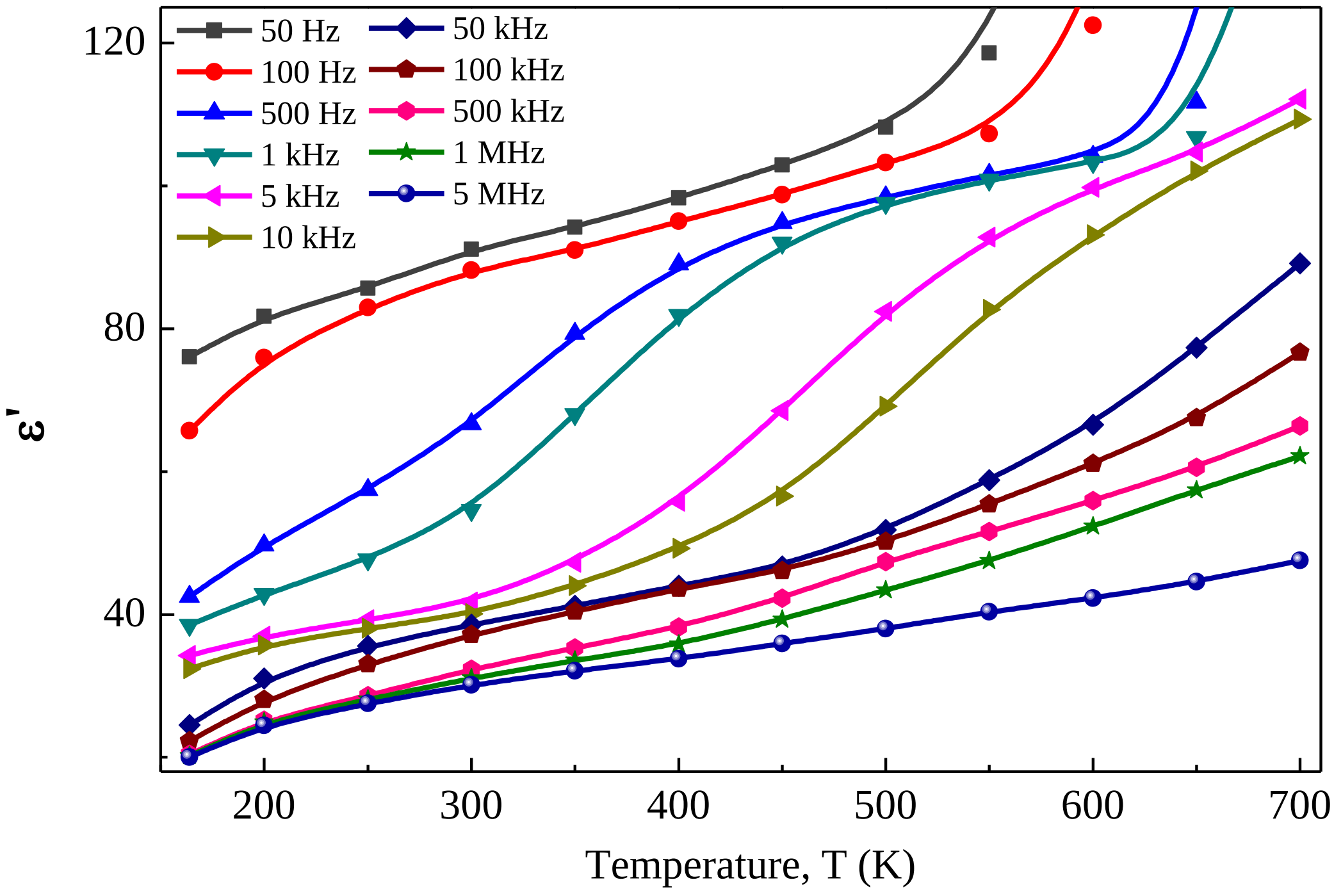
50 Hz
100 Hz
500 Hz
1 kHz
5 kHz
10 kHz
50 kHz
100 kHz
500 kHz
1 MHz
5 MHz
120
80
40
ε'
200
300
400
500
600
700
Temperature, T (K)

**Fig. 9**

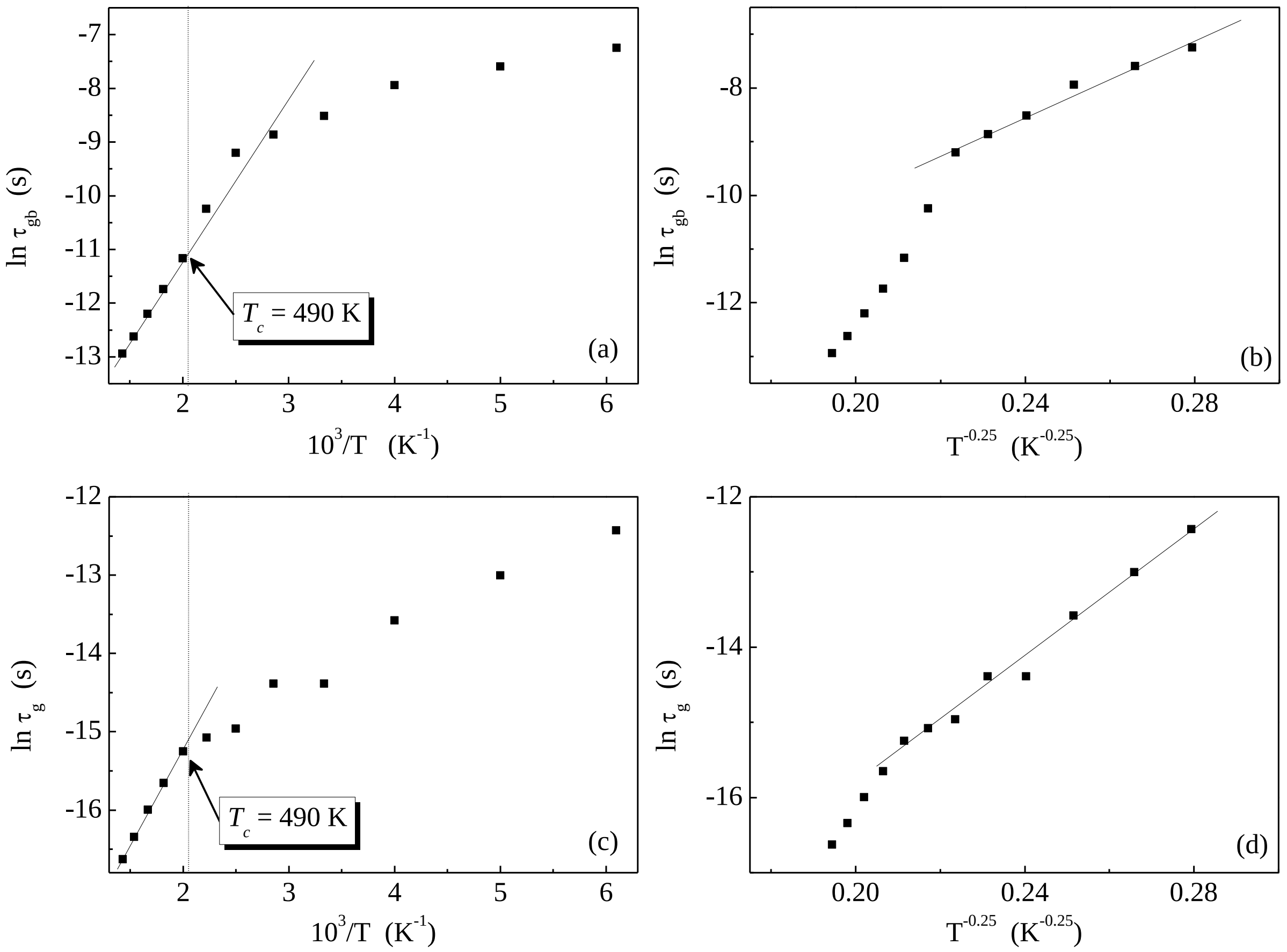

**Fig. 10**

**Figure captions**

Fig. 1: Rietveld refinement results of XRD data for PNT at room temperature. The inset shows the unit cell of PNT as obtained from Rietveld refinement.

Fig. 2: Grain size distribution of the sample PNT calcined at 1573 K. Inset shows the SEM micrograph of PNT samples.

Fig. 3: Raman spectrum of PNT (Experimental data are open circles, while the solid lines represent phonon modes adjusted by Lorentzian curves).

Fig. 4: Representative displacement vectors of $TiO_6$ octahedra. Here red and blue atoms represent oxygen and titanium, respectively.

Fig. 5: Frequency dependence of the ac conductivity (a) at various temperatures, (b) The fitted experimental data with the power law for temperatures 700 K, 500 K, 300 K, and 164 K.

Fig. 6: (a) The thermal variation of dc resistivity for PNT; The dc resistivity of PNT describes the conduction process using (b) small polaron hopping model, (c) Mott's variable range hopping model. The insets show the activation energy (b) and hopping energy (c) as a function of temperature. (d) Variation of hopping range with the temperature.

Fig. 7: Complex impedance plane plots at different temperatures for PNT (a, b, c). Solid lines are fitted data to the observed data (points). (d) The equivalent circuit used to fit the observed data.

Fig. 8: Frequency dependence of $\varepsilon'$ (a) and $\tan\delta$ (c) at various temperatures for PNT. Fitting of $\varepsilon'$ (b) and $\tan\delta$ (d) data using Cole-Cole model.

Fig. 9: Temperature dependence of $\varepsilon'$ at various frequencies for PNT.

Fig. 10: Relaxation times $\tau_{gb}$ and $\tau_g$ of the carriers at grain-boundaries and grains are plotted using (a, c) small polaron hopping model and (b, d) Mott's variable range hopping model.

**Temperature dependent conductivity mechanisms in $Pr_2NiTiO_6$**

***Tables:***

Table 1

| Space group: *P2₁/n* (Monoclinic) | | | | | | |
|---|---|---|---|---|---|---|
| Cell parameters: $a$ = 5.4231 Å, $b$ = 5.4216 Å, $c$ = 7.6829 Å and $\beta$ = 90.116° | | | | | | |
| Reliability factors: $R_{exp}$ = 7.09, $R_p$ = 5.18, $R_{wp}$ = 6.68 and $\chi^2$ = 0.887 | | | | | | |
| Atoms | Wyckoff site | *x* | *y* | *z* | Bond length (Å) | Bond angle (°) |
| Pr | 4*e* | 0.50766 | 0.53271 | 0.24575 | Ni–O1 = 2.0726(4) | Ni–O1–Ti = 155.82(5) |
| Ni | 2*c* | 0.00000 | 0.50000 | 0.00000 | Ni–O2 = 1.9332(4) | Ni–O2–Ti = 150.44(5) |
| Ti | 2*d* | 0.50000 | 0.00000 | 0.00000 | Ni–O3 = 1.8802(3) | Ni–O3–Ti = 131.04(4) |
| O1 | 4*e* | 0.23436 | 0.20455 | -0.04373 | Ti–O1 = 1.8483(4) | |
| O2 | 4*e* | 0.22991 | 0.75693 | 0.06438 | Ti–O2 = 2.0322(4) | |
| O3 | 4*e* | 0.48821 | -0.01115 | 0.25552 | Ti–O3 = 1.9652(4) | |

Table 2

| Band no. | Frequency ($cm^{-1}$) | FWHM ($cm^{-1}$) | Symmetry |
|---|---|---|---|
| 1 | 83 | 32 | *T* |
| 2 | 105 | 17 | *T* |
| 3 | 157 | 30 | *T* |
| 4 | 204 | 18 | *T* |
| 5 | 246 | 43 | *L* |
| 6 | 290 | 10 | *L* |
| 7 | 326 | 25 | *L* |
| 8 | 370 | 19 | *L* |
| 9 | 412 | 35 | $\nu_5$ |
| 10 | 444 | 24 | $\nu_5$ |
| 11 | 465 | 25 | $\nu_5$ |
| 12 | 493 | 32 | $\nu_5$ |
| 13 | 550 | 29 | $\nu_2$ |
| 14 | 592 | 45 | $\nu_2$ |
| 15 | 640 | 55 | $\nu_2$ |
| 16 | 691 | 34 | $\nu_2$ |
| 17 | 742 | 43 | $\nu_1$ |

Table 3

| Temperature (K) | $R_e$ ($\times 10^6$ Ω) | $Q_e$ (*n*F) | $n_e$ | $R_{gb}$ ($\times 10^5$ Ω) | $Q_{gb}$ (*n*F) | $n_{gb}$ | $R_g$ ($\times 10^4$ Ω) | $Q_g$ (*n*F) | $n_g$ |
|---|---|---|---|---|---|---|---|---|---|
| 700 | 6 | 0.08 | 0.99 | 1.0 | 0.015 | 0.99 | 1.2 | 0.005 | 0.92 |
| 500 | 8 | 0.8 | 0.94 | 1.5 | 0.14 | 0.95 | 2.5 | 0.007 | 0.93 |
| 300 | 11 | 4 | 0.92 | 2.1 | 0.8 | 0.93 | 3.9 | 0.02 | 0.94 |
| 164 | 15 | 20 | 0.89 | 3.0 | 3.6 | 0.91 | 5.5 | 0.08 | 0.95 |

Table 4

| Regimes | Temperature (K) | $\Delta\varepsilon = (\varepsilon_S - \varepsilon_\infty)$ | $\omega_m$ (rad/s) | $\alpha$ |
|---|---|---|---|---|
| Electrode contact | | $\Delta\varepsilon_e$ | $(\omega_m)_e$ | $\alpha_e$ |
| | 164 | - | - | - |
| | 300 | 616 | 10 | 0.3 |
| | 500 | 804 | 20 | 0.27 |
| | 700 | 1905 | 1200 | 0.25 |
| Poorly conducting grain boundaries | | $\Delta\varepsilon_{gb}$ | $(\omega_m)_{gb}$ | $\alpha_{gb}$ |
| | 164 | 53 | 5000 | 0.16 |
| | 300 | 55 | 20000 | 0.15 |
| | 500 | 51 | 300000 | 0.1 |
| | 700 | 46 | 2000000 | 0.05 |
| Conducting grains | | $\Delta\varepsilon_g$ | $(\omega_m)_g$ | $\alpha_g$ |
| | 164 | 16 | 1000000 | 0.16 |
| | 300 | 9.5 | 9000000 | 0.14 |
| | 500 | 16 | 30000000 | 0.12 |
| | 700 | 27 | 70000000 | 0.1 |

**Table captions**

Table 1: Structural parameters extracted from the Rietveld refinement of the XRD data for PNT at room temperature.

Table 2: Observed Raman active phonon modes of PNT.

Table 3: Fitted parameters for complex impedance plane plots of PNT.

Table 4: Fitting parameters of $\varepsilon'$ and $\tan\delta$ at various temperatures for PNT as obtained from Cole-Cole equation.